\documentclass[cits]{PoS} 
\title{Decay Constants of Heavy Pseudoscalar Mesons: Reconciling
QCD Sum Rules and Lattice QCD}\ShortTitle{Decay Constants of Heavy
Pseudoscalar Mesons}\author{\speaker{Wolfgang Lucha}\\Institute
for High Energy Physics, Austrian Academy of Sciences,
Nikolsdorfergasse 18, A-1050 Vienna, Austria\\E-mail:
\email{Wolfgang.Lucha@oeaw.ac.at}}\author{Dmitri
Melikhov\\Institute for High Energy Physics, Austrian Academy of
Sciences, Nikolsdorfergasse 18, A-1050 Vienna, Austria,\\Faculty
of Physics, University of Vienna, Boltzmanngasse 5, A-1090 Vienna,
Austria, and\\D.~V.~Skobeltsyn Institute of Nuclear Physics,
Moscow State University, 119991, Moscow, Russia\\E-mail:
\email{dmitri\_melikhov@gmx.de}}\author{Silvano Simula\\INFN,
Sezione di Roma Tre, Via della Vasca Navale 84, I-00146 Roma,
Italy\\E-mail: \email{simula@roma3.infn.it}}

\abstract{Exploiting recently proposed novel improvements of the
techniques used for extracting from QCD sum rules pivotal hadron
characteristics, including their systematic errors, we succeed to
achieve a conspicuous agreement of the predictions of QCD sum
rules (QCD-SR) for the decay constants~$f_B$ and $f_{B_s}$ of the
pseudoscalar mesons $B$ and $B_s,$ which before tended to be
slightly too large, with the corresponding results of lattice-QCD
(LQCD) computations for $N_f$ dynamical sea-quark flavours:
Adopting for the crucial $b$-quark mass
$m_b\equiv\overline{m}_b(\overline{m}_b)$ defined in the
$\overline{\it MS}$ renormalization scheme the value
$m_b=(4.247\pm0.034)\;\mbox{GeV},$ we get
$f_B=(192.0\pm14.6)\;\mbox{MeV}$ and
$f_{B_s}=(228.0\pm19.8)\;\mbox{MeV};$ clearly, this may be also
viewed as a determination of $m_b$ from sufficiently precise decay
constants.

\begin{center}\begin{tabular}{cc}
\includegraphics[scale=.343757]{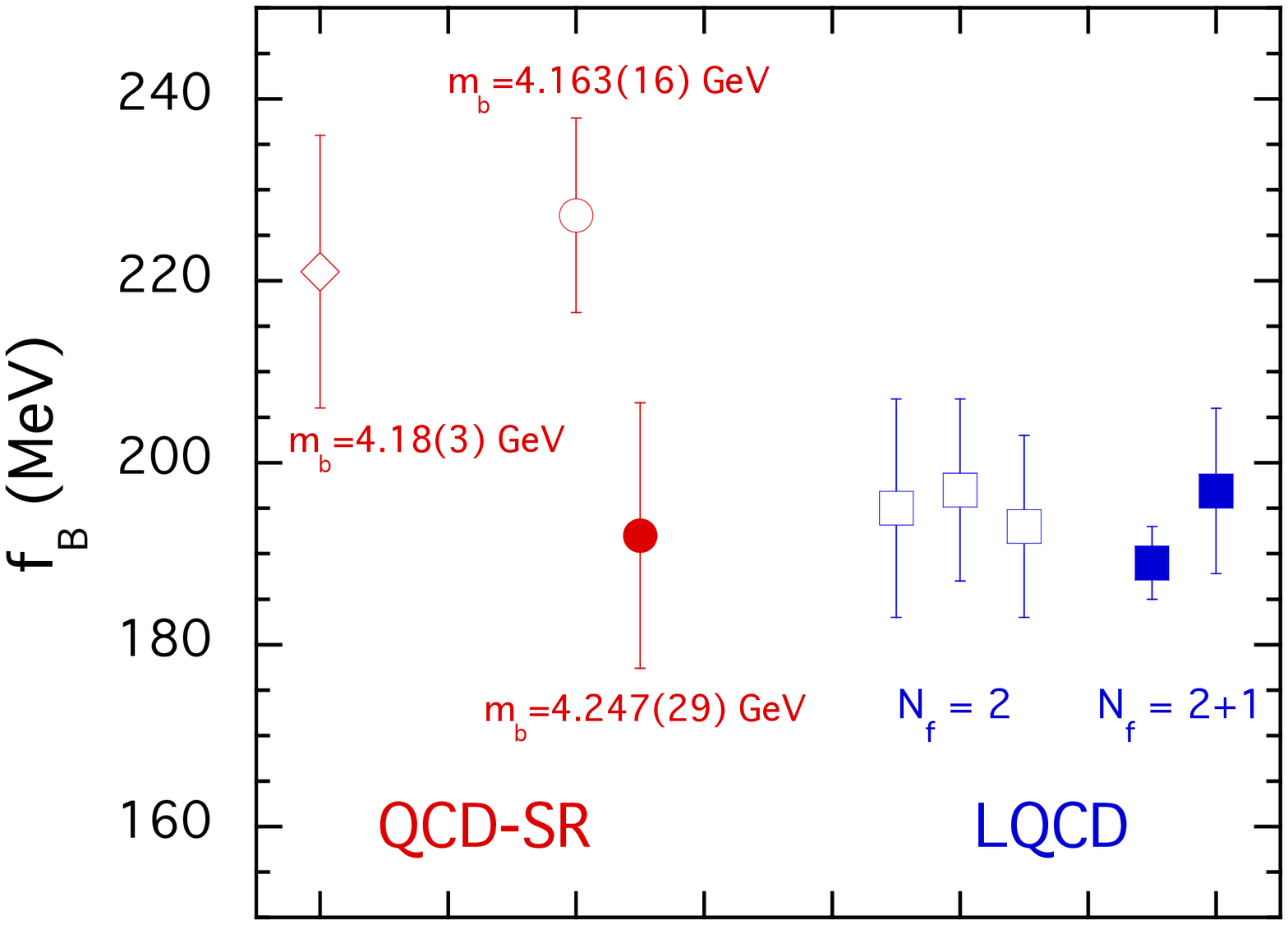}&
\includegraphics[scale=.343757]{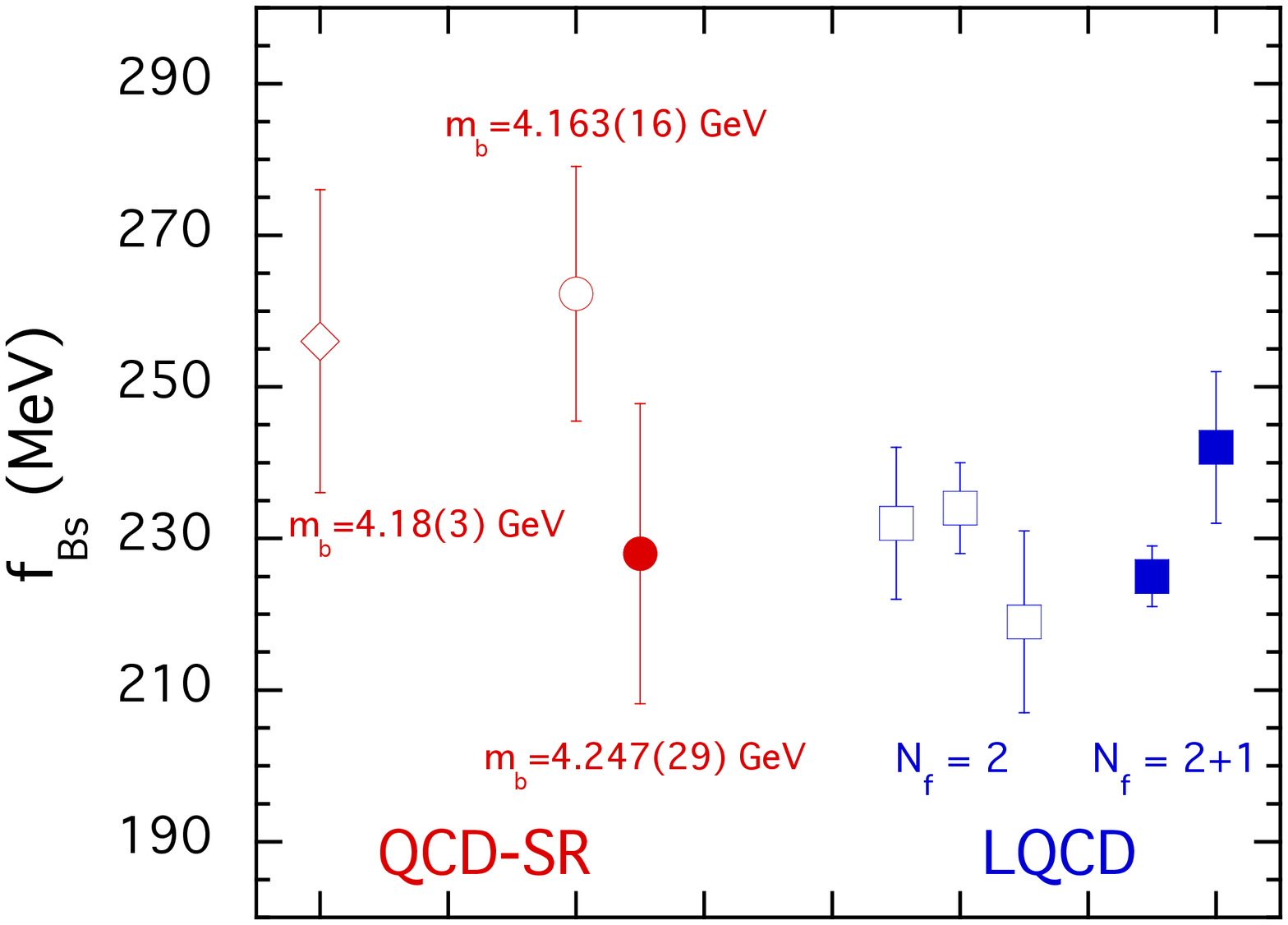}
\end{tabular}\end{center}}

\FullConference{The European Physical Society Conference on High
Energy Physics -- EPS-HEP2013\\18--24 July 2013\\Stockholm,
Sweden}

\begin{document}\section*{Introduction, Cursory Sketch of Basic
Idea, Summary of Findings, and Conclusions}By simultaneous
evaluation of vacuum expectation values of nonlocal products of
interpolating quark currents at both QCD and hadron level, {\em
QCD sum rules\/} relate measurable hadronic features to the
fundamental parameters of QCD \cite{SVZ}. Wilson's operator
product expansion enables us to express any such nonlocal product
as a series of local operators. Our ignorance about higher
resonances can be hidden by assuming {\em quark--hadron
duality\/}: Beyond some effective threshold $s_{\rm eff}$ the
perturbative QCD contributions are expected to counterbalance
those of the hadronic excitations and continuum. Applying {\em
Borel transformations\/} to Borel variables suppresses the impact
of heavier hadronic states.

In order to improve the {\em accuracy\/} of sum-rule predictions
and to estimate reliably the {\em systematic uncertainties\/}, we
recently developed some modifications of the QCD sum-rule method
\cite{LMSAU} centered around the idea to allow $s_{\rm eff}$ to
depend on the Borel variables \cite{LMSET}. This dependence of
$s_{\rm eff}$ is found by minimizing, for polynomial ans\"atze for
$s_{\rm eff},$ the deviation of the predicted meson mass squared
from the true meson mass squared over the range of values allowed
for the Borel variables. For the meson observable of interest, its
systematic error due to the intrinsically limited accuracy of QCD
sum-rule predictions is the spread of results derived for this
observable using $s_{\rm eff}$ ans\"atze~from~linear~to~cubic.

Studying the bottom-meson system along this path \cite{LMSB}, we
note that expressing our sum~rules~in terms of the {\em
$\overline{\it MS}$ running mass\/} of the heavy quark instead of
its {\em pole mass\/} considerably improves the perturbative
convergence of sum-rule findings \cite{JL} and that, contrary to
the charmed-meson case \cite{LMSD}, the decay constants
$f_{B_{(s)}}$ of the $B_{(s)}$ mesons are very sensitive to the
$b$-quark $\overline{\rm MS}$ mass
$m_b\equiv\overline{m}_b(\overline{m}_b).$ Regarding this latter
fact as serendipity, by choosing $m_b=(4.247\pm0.034)\;\mbox{GeV}$
our QCD sum rules entail $f_B=(192.0\pm14.3_{\rm QCD}\pm3.0_{\rm
syst})\;\mbox{MeV}$ and $f_{B_s}=(228.0\pm19.4_{\rm QCD}\pm4_{\rm
syst})\;\mbox{MeV}$ for the decay constants of the $B$ and $B_s$
mesons, which perfectly reproduces our averages
$f_B=(191.5\pm7.3)\;\mbox{MeV}$ and
$f_{B_s}=(228.8\pm6.9)\;\mbox{MeV}$ of several recent lattice-QCD
evaluations \cite{LQCD} of these decay constants.

\vspace{2.6ex}\noindent{\bf Acknowledgments.} D.M.\ was supported
by the Austrian Science Fund (FWF), project no.~P22843.

\end{document}